\documentclass[preprint,apjl,dvipdfmx]{aastex}
\usepackage{natbib}
\shorttitle{Swing Amplification of Spirals}
\shortauthors{Michikoshi \& Kokubo}

\begin{document}

\newcommand{\lamy}{\tilde{\lambda}_y}
\newcommand{\lamymax}{\tilde{\lambda}_{y,\mathrm{max}}}

\title{
  Galactic Spiral Arms by Swing Amplification
}
\author{
  Shugo Michikoshi\altaffilmark{1}, and Eiichiro Kokubo\altaffilmark{1}
}
\altaffiltext{1}{
  Division of Theoretical Astronomy, National Astronomical Observatory of
  Japan, Osawa, Mitaka, Tokyo 181-8588, Japan
}

\email{
  shugo.michikoshi@nao.ac.jp, and kokubo@th.nao.ac.jp
}

\begin{abstract}

Based on the swing amplification model of Julian and Toomre (1966), we
 investigate the formation and structure of stellar spirals in disk
 galaxies.
We calculate the pitch angle, wavelengths, and amplification factor of
 the most amplified mode.
We also obtain the fitting formulae of these quantities as a function of
 the epicycle frequency and Toomre's $Q$.
As the epicycle frequency increases, the pitch angle and radial
 wavelength increases, while the azimuthal wavelength decreases.
The pitch angle and radial wavelength increases with $Q$, while the
 azimuthal wavelength weakly depends on $Q$.
The amplification factor decreases with $Q$ rapidly.
In order to confirm the swing amplification model, we perform local
 $N$-body simulations.
The wavelengths and pitch angle by the swing amplification model are in
 good agreement with those by $N$-body simulations.
The dependence of the amplification factor on the epicycle frequency in
 $N$-body simulations is generally consistent with that in the swing
 amplification model.
Using these results, we estimate the number of spiral arms as a function
 of the shear rate.
The number of spiral arms increases with the shear rate if the disk to
 halo mass ratio is fixed.

\end{abstract}

\section{Introduction}
The origin of spiral structures in disk galaxies is one of the unsolved problems in astrophysics.
The density wave model is a classical model for spiral arm formation \citep{Lin1964, Lin1966}. 
In this model, the spiral arm is considered as a quasi-stationary density wave that rotates around the galactic center with a constant pattern speed.
Since the wave winds much more slowly than the material arm, the winding problem is avoidable.

  Another picture is a dynamic material wave model.
  In a differentially rotating disk, a leading density pattern rotates to a trailing one due to the shear. 
  During the rotation, the leading mode is amplified into spiral arms due to the self-gravity if Toomre's $Q$ is $Q=1 \mbox{--}2$.
  This mechanism is called swing amplification \citep{Goldreich1965, Julian1966, Toomre1981}. 
  In this model, spiral arms are a superposition of many unstable waves.
  The spirals generated by the swing amplification are transient and recurrent, which are observed in $N$-body simulations of multi-arm spirals \citep{Sellwood1984, Baba2009, Sellwood2000, Fujii2011}.

\cite{Julian1966} performed the linear analysis of the collisionless Boltzmann equation and found the swing amplification.
If the spirals are generated by the swing amplification in $N$-body simulations, it is expected that the spirals correspond to the most amplified wave in the linear analysis.
However, \cite{Julian1966} did not investigate the property of the most amplified wave in detail.
\cite{Fuchs2001} investigated the azimuthal wave number of the most amplified wave by the liner analysis of the collisionless Boltzmann equation.
However, his formula is only available for $0.2 < \Gamma < 1.0$, where $\Gamma$ is the shear rate
\begin{equation}
  \Gamma = - \frac{\mathrm{d} \log \Omega}{\mathrm{d} \log r} =  \frac{2 A}{\Omega} = 2 - \frac{\kappa^2}{2\Omega^2},
  \label{eq:gamma}
\end{equation}
where $r$ is the galactocentric distance, $\Omega$ is the circular frequency, $A$ is the Oort constant, and $\kappa$ is the epicycle frequency.
Using this formula, \cite{Baba2013} estimated the pitch angle of spiral arms for $0.2 < \Gamma < 1.0$.

Observationally, it has been known that the pitch angle tends to decrease with $\Gamma$ \citep{Seigar2005, Seigar2006}.
  \cite{Grand2013} performed $N$-body simulations and investigated the pitch angle of spiral arms and found that the galaxies with higher shear rates tend to have smaller pitch angles, which is consistent with the observations.
\cite{Michikoshi2014} (hereafter referred to as Paper I) investigated the pitch angle by local $N$-body simulations and the linear analysis.
We performed local $N$-body simulations with various $\Gamma$ and found that the pitch angle decreases with $\Gamma$.
We obtained the fitting formula of the pitch angle that is available for wider range than that in \cite{Baba2013}.
Based on the linear theory of the swing amplification by \cite{Julian1966}, 
we derived the pitch angle formula for large Toomre's $Q$, which is almost the same as the fitting formula by $N$-body simulations.
This formula also seems consistent with the grand-design spirals in barred galaxies \citep{Baba2015}.
This indicates that the swing amplification can explain some aspects of basic physics of the spiral arm formation.

In this paper, we extend Paper I to investigate the other quantities that characterize spiral arms and understand the swing amplification.
One of the important quantities is the amplification factor, which means how the wave amplitude increases during the swing from leading to trailing.
Using the Goldreich-Lynden-Bell model \citep{Goldreich1965} with the reduction factor \cite{Toomre1981} investigated the dependence of the amplification factor on the azimuthal wavelength for $\Gamma=1$.
\cite{Athanassoula1984} and \cite{Dobbs2014} performed the same analysis for general $\Gamma$.
However, in these works, the most amplified wave was not focused on. 
In Paper I, we showed that the pitch angle of spiral arms in $N$-body simulations can be explained by the most amplified wave in the linear theory.
Thus, it is important to investigate the most amplified wave.
The dependence of the wavelength and amplification factor of the most amplified wave on $\Gamma$ and $Q$ has not been investigated by the linear theory in detail yet.
Moreover, the property of the most amplified wave has not been confirmed by $N$-body simulations quantitatively.

The goal of this paper is to extend Paper I to calculate the amplification factor and radial and azimuthal wavelengths of the most amplified wave, 
which enables us to understand the structures of spiral arms. 
The outline of this paper is as follows.
Section 2 deals with the Julian-Toomre model and gives the wavelength of the most amplified wave and the amplification factor.
In section 3, we compare the results of the Julian-Toomre model with those by local $N$-body simulations.
Section 4 is devoted to a summary and discussions.

\section{Linear Analysis \label{sec:jt}}
\subsection{Integral Equation}
First we briefly summarize the Julian-Toomre model \citep{Julian1966}.
We consider local Cartesian coordinates $(x,y,z)$ whose origin revolves around the galactic center with the circular frequency $\Omega$.
The $x$- and $y$-axes are directed along the radial and rotational directions, respectively, and 
the $z$-axis is normal to the $x$-$y$ plane.

We consider the plane wave with the radial wavelength $\lambda_x$ and the azimuthal wavelength $\lambda_y$.
Because of the shear, the wave rotates from a leading wave to a trailing one.
While $\lambda_x$ changes with time $t$, $\lambda_y$ remains constant.
We set $t=0$ when the radial wavenumber $k_x=0$, that is, the pitch angle is $90^\circ$.
Thus, the negative time $t<0$ corresponds to a leading wave, and the positive time corresponds to a trailing one.
We use the non-dimensional parameter $\lamy=\lambda_y/\lambda_\mathrm{cr}$ where $\lambda_\mathrm{cr} = 4 \pi^2 G \Sigma_0/\kappa^2$ and $\Sigma_0$ is the unperturbed surface density, which is the same as $X$ in \cite{Julian1966}.
During the rotation, the wave amplitude changes.

\cite{Julian1966} derived the equation for the evolution of the density amplitude from the collisionless Boltzmann equation in the corotating frame of the wave. 
The time evolution of the density $D$ with the imposed density perturbation $D_\mathrm{imp}$ is described by the integral equation
\begin{equation}
  D(t, \kappa, Q, t_\mathrm{i}, \lamy) = \int_{t_\mathrm{i}}^t K(t',t, \kappa, Q, \lamy) (D_\mathrm{imp} + D(t')) \mathrm{d}t', 
  \label{eq:int0}
\end{equation}
where $t_\mathrm{i}$ is the initial time and $K$ is the kernel function 
\begin{equation}
  K(t',t, \kappa, Q, \lamy) = \frac{0.32 \pi}{\lamy} \frac{C_\mathrm{c}(t')C_\mathrm{d}(t) - C_\mathrm{d}(t') C_\mathrm{c}(t)}{\sqrt{1+4 A^2 t'^2}}
  \exp \left( - 0.143 \frac{Q^2}{\lamy^2} ((C_\mathrm{c}(t') -C_\mathrm{c}(t))^2 + (C_\mathrm{d}(t') - C_\mathrm{d}(t))^2)  \right),
\end{equation}
\begin{equation}
  C_\mathrm{c}(t) = 2 A t \cos \kappa t  - \frac{2 \Omega}{\kappa} \sin \kappa t,
\end{equation}
\begin{equation}
  C_\mathrm{d}(t) = 2 A t \sin \kappa t  + \frac{2 \Omega}{\kappa} \cos \kappa t,
\end{equation}
and $Q=\sigma_x \kappa/(3.36 G \Sigma_0)$ is Toomre's $Q$ and $\sigma_x$ is the radial velocity dispersion \citep{Toomre1964}.
As seen in Equation (2), $D$ is a function of two disk parameters, $\kappa$ and $Q$, and two wave parameters $t_\mathrm{i}$ and $\lamy$.
Assuming that the imposed density is described by the delta function $D_\mathrm{imp} = \delta(t-t_\mathrm{i})$ and using the Simpson's rule, we solve Equation (\ref{eq:int0}) numerically.

\subsection{Most Amplified Wave}

\begin{figure}
  \plottwo{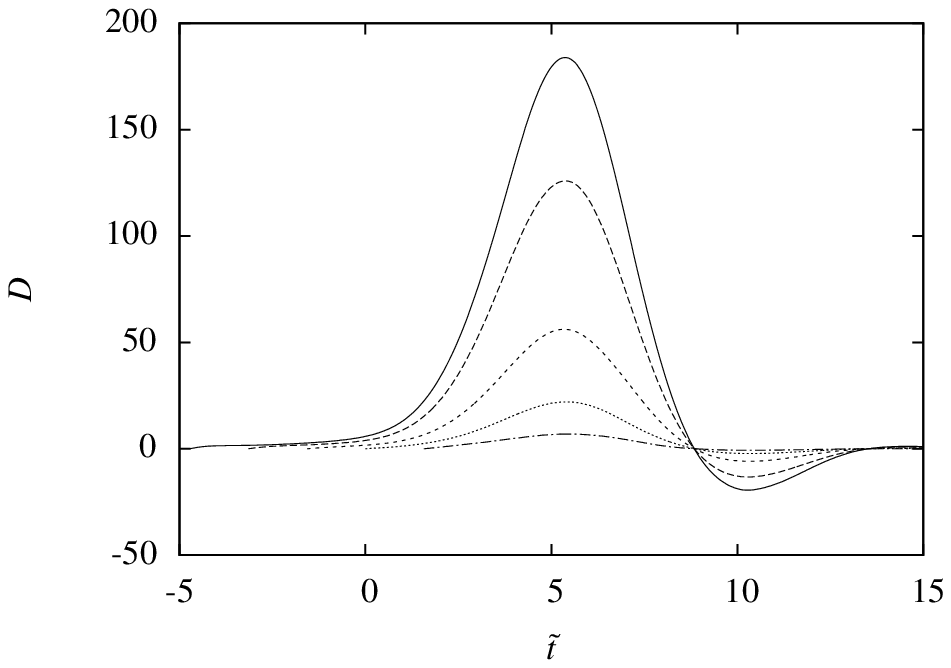}{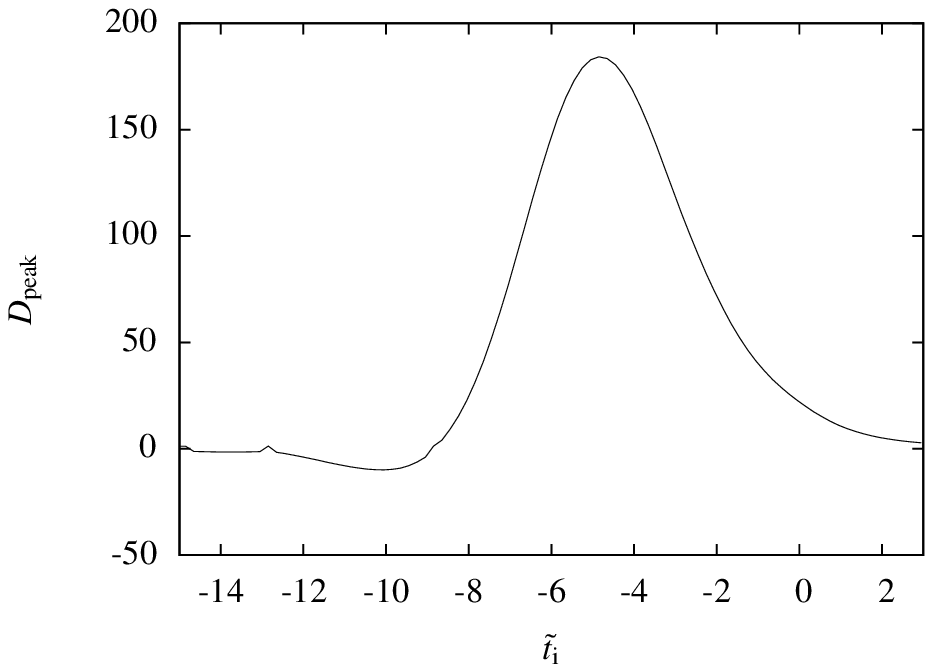}
  \caption{Time evolution of $D$ (left panel) and the peak density amplitude $D_\mathrm{peak}$ as a function of $t_\mathrm{i}$ (right panel).
	The initial time in the left panel is $\tilde t_\mathrm{i} = -1.5 \pi$ (solid), $  - 1.0 \pi$ (dashed), $ - 0.5 \pi$ (short dashed), $ 0.0 \pi$ (dotted), and $ 0.5 \pi$ (dotted dashed). The other parameters are $\kappa/\Omega=\sqrt{2}$, $Q=1.0$, and $\lamy=1.0$. 
This Figure is the same as Fig. $4$ in \cite{Julian1966}.
}
	\label{fig:den_evo_jt}
\end{figure}

The left panel of Figure \ref{fig:den_evo_jt} shows the time evolution of $D$ with $\kappa/\Omega =\sqrt{2}$, $Q=1$, $\tilde t_\mathrm{i} = -1.5 \pi, -1.0\pi, -0.5\pi, 0, 0.5\pi$, and $\lamy=1$. 
The variable $\tilde t$ is the normalized time $\tilde t = \kappa t$.
The density amplitude has a peak at the positive time, that is, the most amplified wave is trailing.
The peak density $D_\mathrm{peak}$ depends on the initial time $\tilde t_\mathrm{i} = \kappa t_\mathrm{i}$, but the time for $D_\mathrm{peak}$ barely depends on $\tilde t_\mathrm{i}$, which is $\tilde t \simeq 5.3$.
Optimizing $\tilde t_\mathrm{i}$, we can calculate the maximum amplitude with fixed $\lamy$. 
The right panel of Figure \ref{fig:den_evo_jt} shows the dependence of $D_\mathrm{peak}$ on $\tilde t_\mathrm{i}$.
The wave is most amplified with $\tilde t _\mathrm{i}=-4.85$, where $D_\mathrm{peak}$ is $184$. 

\begin{figure}
  \plotone{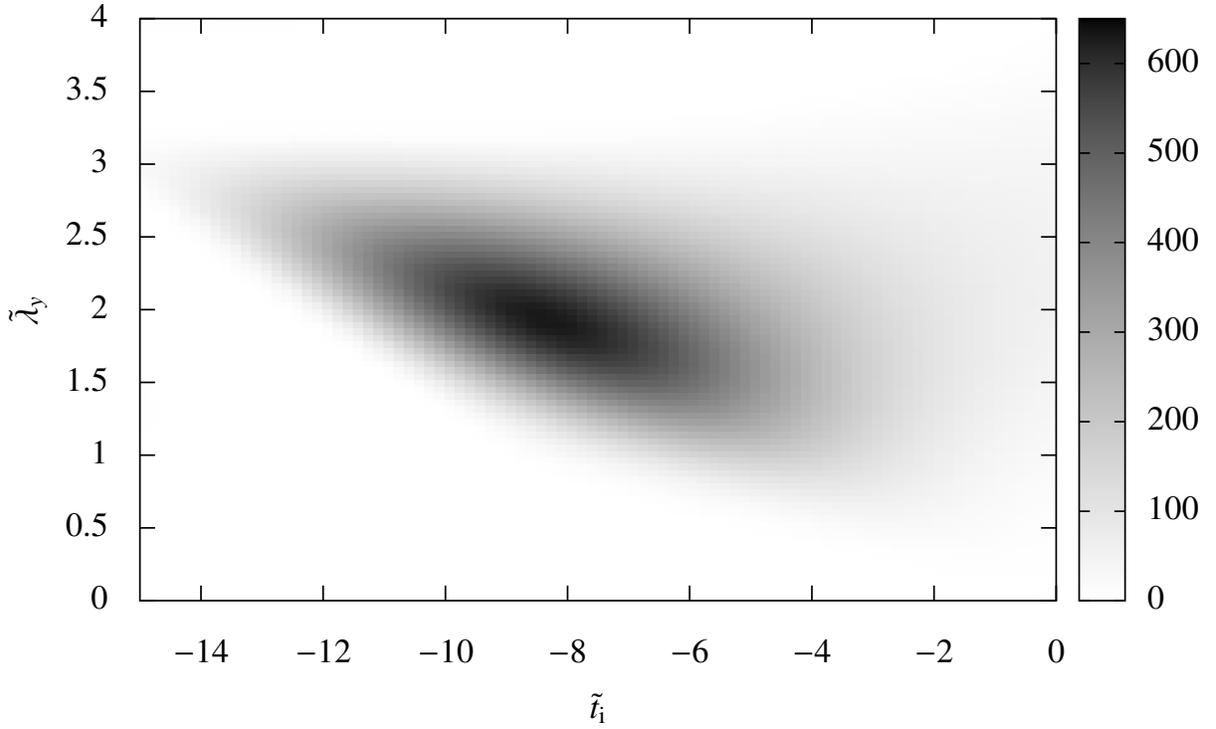}
  \caption{Amplification factor $D$ as a function of $\tilde t_\mathrm{i}$ and $\lamy$ for $\kappa/\Omega=\sqrt{2}$ and $Q=1$.}
	\label{fig:t0xdep}
\end{figure}

More generally, $D_\mathrm{peak}$ depends on $\tilde t_\mathrm{i}$ and $\lamy$. 
Thus, by optimizing $\tilde t_\mathrm{i}$ and $\lamy$ we can calculate the most amplified wave for the disk model given by $\kappa$ and $Q$.
Figure \ref{fig:t0xdep} shows $D_\mathrm{peak}$ for various $t_\mathrm{i}$ and $\lamy$ for $\kappa/\Omega =\sqrt{2}$ and $Q=1$.
The peak density has the maximum value $D_\mathrm{max}$ for $\tilde t_\mathrm{i}=-8.2$ and $\lamy=1.9$.
This most amplified wave is considered as the wave observed in $N$-body simulation.

\subsection{Fitting Formula}
We investigate the pitch angle $\theta_\mathrm{max}$, the azimuthal wavelength $\lamymax$, the radial wavelength $\tilde \lambda_{x,\mathrm{max}}$, and the amplification factor $D_\mathrm{max}$ of the most amplified wave, where $\tilde \lambda_{x,\mathrm{max}} =  \lambda_{x,\mathrm{max}}/ \lambda_{\mathrm{cr}}$ is calculated by
\begin{equation}
  \tilde \lambda_{x,\mathrm{max}} = \lamymax \tan \theta_\mathrm{max}.
  \label{eq:lambdafor}
\end{equation}
Figure \ref{fig:pitchjt} shows the dependence of these quantities on $\kappa/\Omega$ with $Q=1.0, 1.4,$ and $1.8$.
We obtain their fitting formulae as a function of $\kappa/\Omega$ and $Q$.

\subsubsection{Pitch Angle}
We have already obtained the fitting formula of the pitch angle in Paper I
\begin{equation}
  \tan \theta_\mathrm{max} = \frac{\kappa}{7 A},
	\label{eq:fit_pitch}
\end{equation}
where we neglected the weak dependence on $Q$ considering the $N$-body simulation results.
In this paper, we derive a more general fitting formula based on the linear theory.

The most amplified wave has the maximum amplitude at time $\tilde t_\mathrm{max}$.
The corresponding pitch angle $\theta_\mathrm{max}$ of the most amplified wave is
\begin{equation}
  \tan \theta_\mathrm{max} = \frac{\kappa}{2 A \tilde t_{\mathrm{max}}}.
\label{eq:pitch00}
\end{equation}
According to Paper I, $\tilde t_\mathrm{max}$ decreases with $Q$ and approaches $\sim 3$ for large $Q$.
Considering these behaviors, we assume 
\begin{equation}
  \tilde t_\mathrm{max} =\pi\left(1+ \frac{\alpha}{Q^\beta} \right).
  \label{eq:taumax}
\end{equation}
Substituting Equation (\ref{eq:taumax}) into Equation (\ref{eq:pitch00}) , we obtain the pitch angle 
\begin{equation}
  \tan \theta_\mathrm{max} = \frac{1}{2 \pi} \left(1+ \frac{\alpha}{Q^\beta} \right)^{-1}\frac{\kappa}{A}.
	\label{eq:fit_pitch2}
\end{equation}
Using the least square fit method, we obtain $\alpha=2.095$ and $\beta=5.30$.
The relative error is less than 4.7 \% for $Q \ge 1.5$, while the maximum relative error is about 30 \% for $Q \le 1.4$.

In practice, the range $1.5 \le Q \le 2.0$ is important.
For example, in Paper I, regardless of the initial $Q$, $Q$ quickly increases and becomes larger than about $1.5$ due to gravitational scattering by the spiral structure.
In this range, $1/(2 \tilde t_\mathrm{max})$ increases with $Q$ from $0.13$ to $0.15$, indicating that the dependence on $Q$ is weak.
Thus, setting $1/(2 \tilde t_\mathrm{max})$ as the intermediate value $0.14$, we obtain the almost same formula as Equation (\ref{eq:fit_pitch}).

\subsubsection{Azimuthal Wavelength}

\begin{figure}
  \begin{center}
  	\includegraphics[width = 0.6\textwidth] {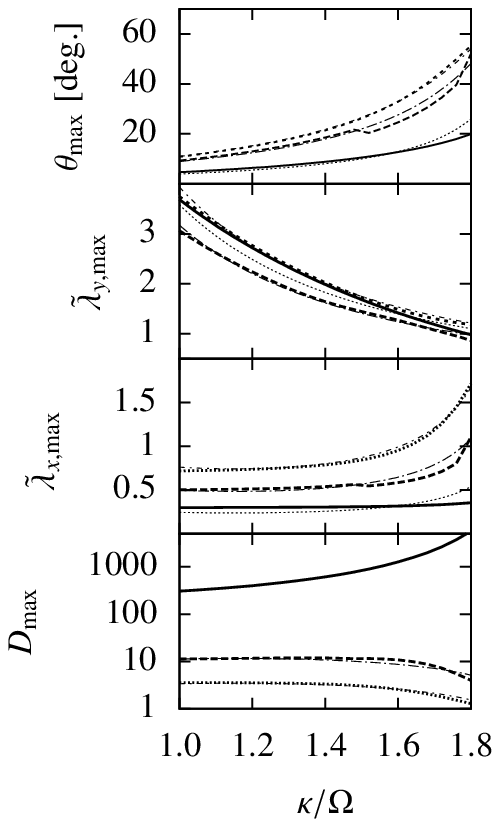}
  \end{center}
  \caption{Spiral parameters $\theta_\mathrm{max}$, $\lamymax$, $\tilde \lambda_{x,\mathrm{max}}$ and $D_\mathrm{max}$ as a function of $\kappa / \Omega$ for $Q=1.0$ (solid curve), $Q=1.4$ (dashed curve), and $Q=1.8$ (short dashed curve). 
	The dotted ($Q=1.0$), dashed-dotted ($Q=1.4$), and short dashed-dotted  ($Q=1.8$) curves denote the fitting formulae given by Equations (\ref{eq:fit_pitch2}), (\ref{eq:fit_xf2}), (\ref{eq:fit_xl}), and (\ref{eq:fit_amp}), respectively.
\label{fig:pitchjt}
}
\end{figure}

\cite{Toomre1981} concluded that $\lamymax $ for the efficient swing amplification is $\lamymax =1 \mbox{--} 2$.
As shown in Figure \ref{fig:pitchjt}, the amplification is significant for $\lamymax =1\mbox{--}4$ and decreases with $\kappa /\Omega$.
We find that the simple power law $\lamymax \propto (\kappa/\Omega)^{-2}$ is appropriate for any $Q$ and $\lamymax$ weakly depends on $Q$.
While $\lamymax $ decreases with $Q$ for $Q \lesssim 1.4$, $\lamymax$ increases with $Q$ for $Q \gtrsim 1.4$.
Assuming $\lamymax = (C_1 Q^2 + C_2 Q + C_3)(\Omega/\kappa)^2$ and using the least square fit method, we obtain the following formula 
\begin{equation}
  \lamymax  = (3.653 Q^2 -9.789Q+9.721) \left(\frac{\Omega}{\kappa}\right)^2.
	\label{eq:fit_xf2} 
\end{equation}
The relative error of this formula is less than 20 \%.
For the range $1.4 \le Q \le 1.8$, we obtain the simpler formula
\begin{equation}
\lamymax =  2.17 Q \left(\frac{\Omega}{\kappa}\right)^2,
\label{eq:fit_xf_simp}
\end{equation}
where the relative error is less than 9 \%.

\subsubsection{Radial Wavelength}

Substituting Equations (\ref{eq:fit_pitch2}) and (\ref{eq:fit_xf2}) into Equation (\ref{eq:lambdafor}), we obtain the radial wavelength
\begin{equation}
  \tilde \lambda_{x,\mathrm{max}}  = \frac{0.581Q^2-1.558Q+1.547}{1+2.095Q^{-5.3}} \frac{\Omega^2}{A \kappa}.
	\label{eq:fit_xl} 
\end{equation}
As already mentioned, for $1.5 \le Q \le 2.0$, we can approximate $ 1/(2 \tilde t_\mathrm{max}) \simeq 0.14$.
In this case, using Equation (\ref{eq:fit_xf_simp}), we obtain the simpler fitting formula
\begin{equation}
  \tilde \lambda_{x,\mathrm{max}} = 0.304 Q \frac{\Omega^2}{A \kappa}.
	\label{eq:fit_xl2} 
\end{equation}
This formula agrees well with the linear analysis. 
For example, $\tilde \lambda_{x,\mathrm{max}}$ is almost constant for $\kappa/\Omega< 1.5$ and is about $0.7$ for $Q=1.8$, and
for $\kappa/\Omega> 1.5$, $\tilde \lambda_{x,\mathrm{max}} $ increases and exceeds $1.0$.

\subsubsection{Amplification Factor}

The maximum amplification factor $D_\mathrm{max}$ increases with decreasing $Q$.
This is because small $Q$ means that the self-gravity is strong relative to the stabilizing effects due to the rotation and the velocity dispersion.
The dependence on $\kappa/\Omega$  is somewhat complicated.
For $Q=1.0$, $D_\mathrm{max}$ increases with $\kappa/\Omega$ from $308$ to $5697$.
On the other hand, for $Q=1.4$, $D_\mathrm{max}$ is almost constant for $\kappa/\Omega<1.5$ and is about $10$ and decreases with $\kappa/\Omega$ for $1.6<\kappa/\Omega$.
Similarly for $Q=1.8$, $D_\mathrm{max}$ is almost constant for $\kappa/\Omega<1.5$ is about $3.5$ and decreases with $\kappa/\Omega$ for $1.6<\kappa/\Omega$.
For $Q > 1.4$, we find that the dependence of $D_\mathrm{max}$ on $\kappa/\Omega$ is roughly described by $\kappa A / \Omega^2$.
Considering various function forms, we find that the dependence on $Q$ is approximated by $\propto \exp(C_\mathrm{q}/Q)$ where $C_\mathrm{q}$ is a constant.
Using the least square fit method, we obtain $D_\mathrm{max}$ for $Q>1.4$ as
\begin{equation}
D_\mathrm{max} = 0.0657 \exp\left( \frac{7.61}{Q} \right) \frac{\kappa A}{\Omega^2}.
	\label{eq:fit_amp} 
\end{equation}
This fitting formula has relatively large error, but it roughly agrees with the linear analysis,
which demonstrates the sensitive dependence on $Q$.

\section{$N$-Body Simulation}
We perform local $N$-body simulations of pure stellar disks based on the epicycle approximation.
The calculation method is the same as that in Paper I.
We briefly summarize the method.

\subsection{Model}

 We consider a small patch of a disk such that $L_x, L_y \ll r$, where $L_x$ and $L_y$ are the width and length of the patch, and $r$ is the galactocentric distance.
 We adopt a local Cartesian coordinate system ($x,y,z$) that is the same as Section 2.
In the epicycle approximation, the equation of motion of particle $i$ is given by 
\begin{eqnarray}
  \frac{\mathrm d^2 x_i}{\mathrm{d}t^2} &=& 2 \Omega \frac{\mathrm{d} y_i}{\mathrm{d}t} + \left(4 \Omega^2 - \kappa^2 \right) x_i + \sum_{j \ne i} \frac{G m (x_j - x_i)}{(r_{ij}^2+\epsilon^2)^{3/2}}, \nonumber  \\
  \frac{\mathrm d^2 y_i}{\mathrm{d}t^2} &=& - 2 \Omega \frac{\mathrm d x_i}{\mathrm{d}t} + \sum_{j \ne i} \frac{G m (y_j - y_i)}{(r_{ij}^2+\epsilon^2)^{3/2}},  \label{eq:eom} \\
  \frac{\mathrm d^2 z_i}{\mathrm{d}t^2} &=& - \nu^2 z_i + \sum_{j \ne i} \frac{G m (z_j - z_i)}{(r_{ij}^2+\epsilon^2)^{3/2}}, \nonumber
\end{eqnarray}
 where $r_{ij}$ is the distance between particles $i$ and $j$, $m$ is the particle mass
 \citep[e.g.,][]{Toomre1981, Toomre1991, Kokubo1992, Fuchs2005}. 
We assume that all particles have the same mass. 
In Equation (\ref{eq:eom}), $2 \Omega \mathrm{d} y_i/\mathrm{d}t$ and $- 2 \Omega \mathrm d x_i/\mathrm{d}t $ are the Coriolis force, $4 \Omega^2 x_i$ is the centrifugal force, $-\kappa^2 x_i$ and $-\nu^2 z_i$ are the galactic gravitational force, and the terms proportional to $(r^2_{ij} + \epsilon^2)^{-3/2}$ are the gravitational force from the other particles.
The length $\epsilon$ is the softening parameter $\epsilon = r_\mathrm{t}/4$ where $r_\mathrm{t}$ is the tidal radius of a particle.
The frequencies $\kappa$ and $\nu$ are the epicycle and vertical frequencies.
We calculate the motion of particles only in the computational box considering the periodic boundary condition \citep[e.g.,][Paper I]{Wisdom1988, Toomre1991, Fuchs2005}.

The size of the computational box $L_x$ and $L_y$ should be sufficiently larger than $\lambda_\mathrm{cr}$.
We set the size of the computational box as $L_x = L_y = L = 15 \lambda_\mathrm{cr}$.
We set the unit time as $\Omega^{-1}$ and the unit length as $r_\mathrm{t}$ \citep{Kokubo1992}.
The equation of motion is integrated using a second-order leapfrog integrator with time-step $\Delta t = (2 \pi / \Omega)/200$. 
We calculate the self-gravity of particles considering the periodic boundary condition.
The cutoff length of the gravity is $L_\mathrm{cut} = L$.
The self-gravity of particles, which is the most computationally expensive part, is calculated using the special-purpose computer, GRAPE-7 \citep{Kawai2006}.

\subsection{Initial Conditions}

We assume that the initial surface density $\Sigma_0$ of particles in the computational box is uniform.
The total mass in area $\lambda_\mathrm{cr}^2$ is fixed and the
 particle mass is given by 
 $m = \lambda_\mathrm{cr}^2 \Sigma_0/N_\mathrm{c}$ 
 where $N_\mathrm{c}$ is the number of particles in $\lambda_\mathrm{cr}^2$. 
We set $N_\mathrm{c}=4000$ and then the total number of particles is
$N=N_\mathrm{c} L_x L_y/\lambda_\mathrm{cr}^2=9 \times 10^5$. 
The number of particles is sufficiently large (Paper I).
Thus the two-body relaxation barely affects the dynamical evolution.

The initial Toomre's $Q$ is 
\begin{equation}
  Q_\mathrm{ini}= \frac{\sigma_x \kappa}{3.36 G \Sigma_0},
  \label{eq:qdef}
\end{equation}
where $\sigma_x$ is the initial radial velocity dispersion \citep{Toomre1964}.  
We vary $\kappa/\Omega$ in $1.0 \le \kappa/\Omega \le 1.8$ ($1.5 \ge \Gamma \ge 0.38$) with fixed $Q_\mathrm{ini}=1.4$ and vary $Q_\mathrm{ini}$ in $1.2 \le Q_\mathrm{ini} \le 1.8$ with fixed $\kappa/\Omega=1.4$ ($\Gamma=1.02$).
The initial radial velocity dispersion $\sigma_x$ is calculated from Equation (\ref{eq:qdef}) using $Q_\mathrm{ini}$.
We adopt the triaxial Gaussian model as the initial velocity distribution (Paper I).
The vertical distribution of particles is determined so that it is consistent with the velocity distribution, and $x$ and $y$ of particles are distributed randomly.
The initial disk parameters are summarized in Table 1.
We performed the simulations with smaller $N_\mathrm{c}$ and $L$ such as $N_\mathrm{c}=2000$ and $L=10\lambda_\mathrm{cr}$, and confirmed that the following results barely depend on $N_\mathrm{c}$ and $L$.

\subsection{Formation of Spiral Arms \label{sec:form_spiral}}
\begin{figure}
 \begin{minipage}{0.49\hsize}
 \begin{center} (A) $t \Omega = 0.0$ \end{center}
   \includegraphics[width=\columnwidth]{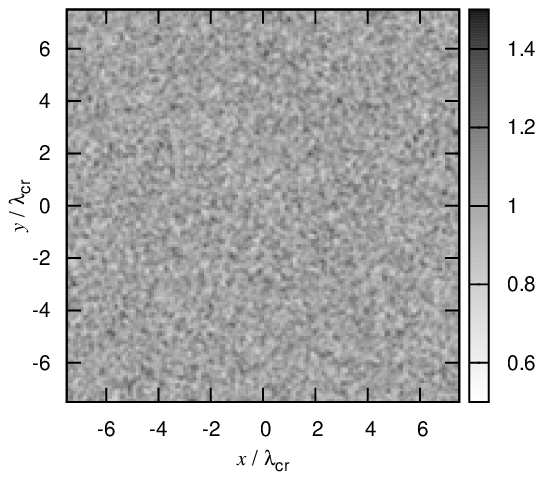}
 \end{minipage}
 \begin{minipage}{0.49\hsize}
 \begin{center} (B) $t \Omega = 0.5$ \end{center}
   \includegraphics[width=\columnwidth]{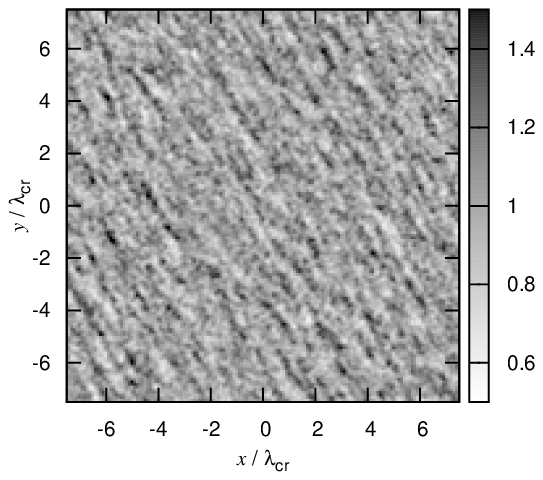}
 \end{minipage}
 \begin{minipage}{0.49\hsize}
 \begin{center}  (C) $t \Omega = 1.0$ \end{center}
   \includegraphics[width=\columnwidth]{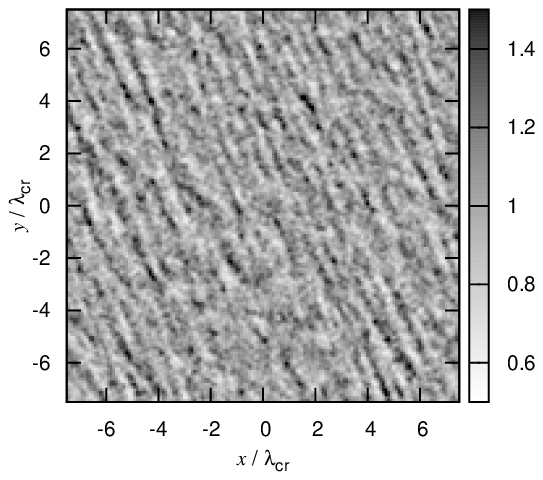}
 \end{minipage}
 \begin{minipage}{0.49\hsize}
 \begin{center}  (D) $t \Omega = 1.5$ \end{center}
   \includegraphics[width=\columnwidth]{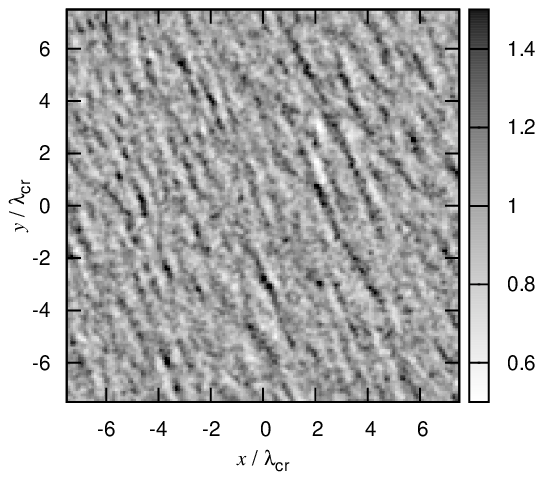}
 \end{minipage}
 \caption{Snapshots of the surface density at $t \Omega = 0.0$, $0.5$, $1.0$, and $1.5$ for $\kappa/\Omega=1.4$ (model k4).}
  \label{fig:initial_snapshot}
\end{figure}

Figure \ref{fig:initial_snapshot} presents the initial evolution of the surface density for model k4.
At $t \Omega=0$, the particles are distributed uniformly and there are not any structures.
Immediately, at $t \Omega=0.5$, the spiral structure is generated spontaneously.
At $t \Omega=1.0, 1.5$, we can observe clear spirals and there is no difference between the two snapshots.
These spirals are not steady but transient and recurrent, which are generated and destroyed continuously.
This activity continues throughout the simulation time.

\begin{figure}
  \begin{center}
  	\includegraphics[width = 0.7\textwidth] {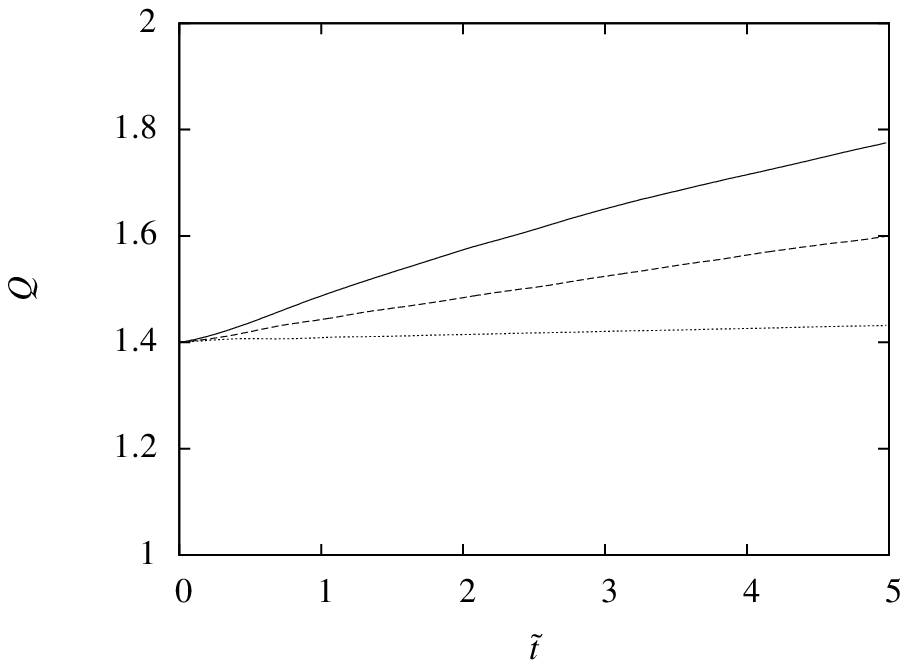}
  \end{center}
  \caption{Time evolution of $Q$ for $\kappa/\Omega=1.0$ (solid), $1.4$ (dashed), and $1.8$ (dotted) (models k0, k4, and k8).}
  \label{fig:tevo}
\end{figure}

Figure \ref{fig:tevo} presents the time evolution of $Q$.
The particles can be heated by particle-particle interactions or particle-spiral interactions.
As mentioned above, the two-body relaxation time is sufficiently longer than the simulation time, which means that particle-particle interactions are negligible.
Thus, the particles are mainly heated by the spiral arms.
As we do not consider any cooling processes in the $N$-body simulations, $Q$ increases monotonically. 
With the larger amplitude of spiral arms, heating is more efficient.
Since the amplitude of spiral arms in the disk with smaller $\kappa/\Omega$ is larger, $Q$ increases more quickly.
In any models, during the simulation time, $Q$ is $1.4 \lesssim Q \lesssim 1.8$.

\subsection{Extraction of Dominant Wave}
\begin{figure}
  \begin{center}
  	\plotone {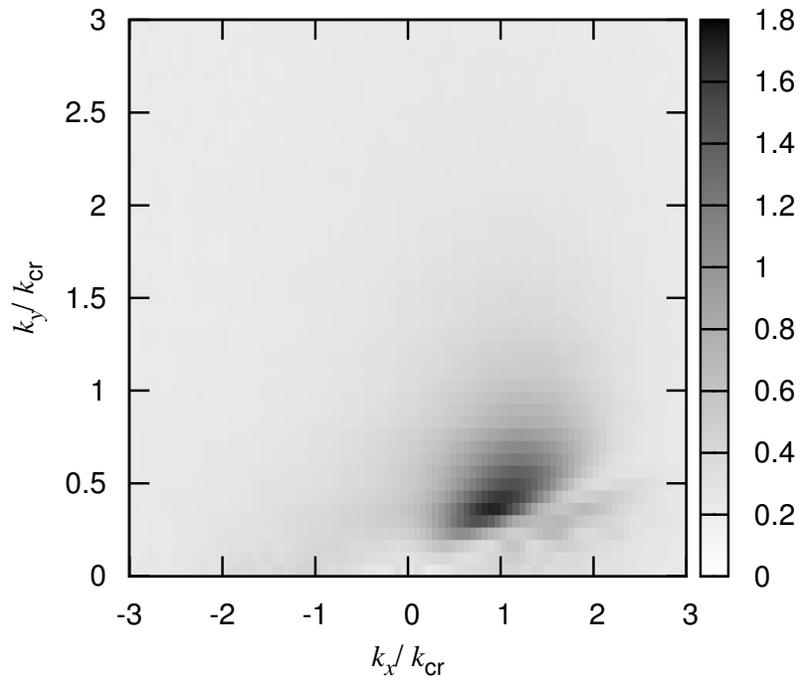}
  \end{center}
  \caption{Time averaged Fourier amplitude $\delta$ for $\kappa/\Omega=1.4$ (model k4). }
  \label{fig:fft}
\end{figure}

In order to extract a dominant wave, we adopt two methods: a Fourier transformation method and a spatial correlation method. 
The Fourier amplitude of the density fluctuation is defined by 
\begin{equation}
  a(k_x, k_y) = \int \!\!\! \int \left( \frac{\Sigma(x,y)}{\Sigma_0}-1 \right) \exp(i (k_x x + k_y y)) \mathrm{d}x \mathrm{d}y,
\end{equation}
where $k_x$ and $k_y$ are the radial and azimuthal wavenumbers.
We calculate the time average of the Fourier amplitude over $t \Omega = 1 \mbox{--} 5$ and find the wavenumber that maximizes the time averaged $\delta$, which corresponds to the dominant wave.
The result is shown in Figure \ref{fig:fft}.
In model k4, the Fourier amplitude has a peak at $k_{x,\mathrm{max}} / k_\mathrm{cr} = 1.0$ and $k_{y,\mathrm{max}} / k_\mathrm{cr}= 0.4$ where $k_\mathrm{cr}=2\pi/\lambda_\mathrm{cr}$.
The corresponding wavelengths are $\tilde \lambda_{x,\mathrm{max}} = 1.00$ and $\lamymax= 2.50$.
The pitch angle is calculated from $\tan \theta_\mathrm{max} = k_{y,\mathrm{max}}/k_{x,\mathrm{max}}$, which is $\theta_\mathrm{max} \simeq 21^\circ$.

\begin{figure}
  \begin{center}
  	\plotone {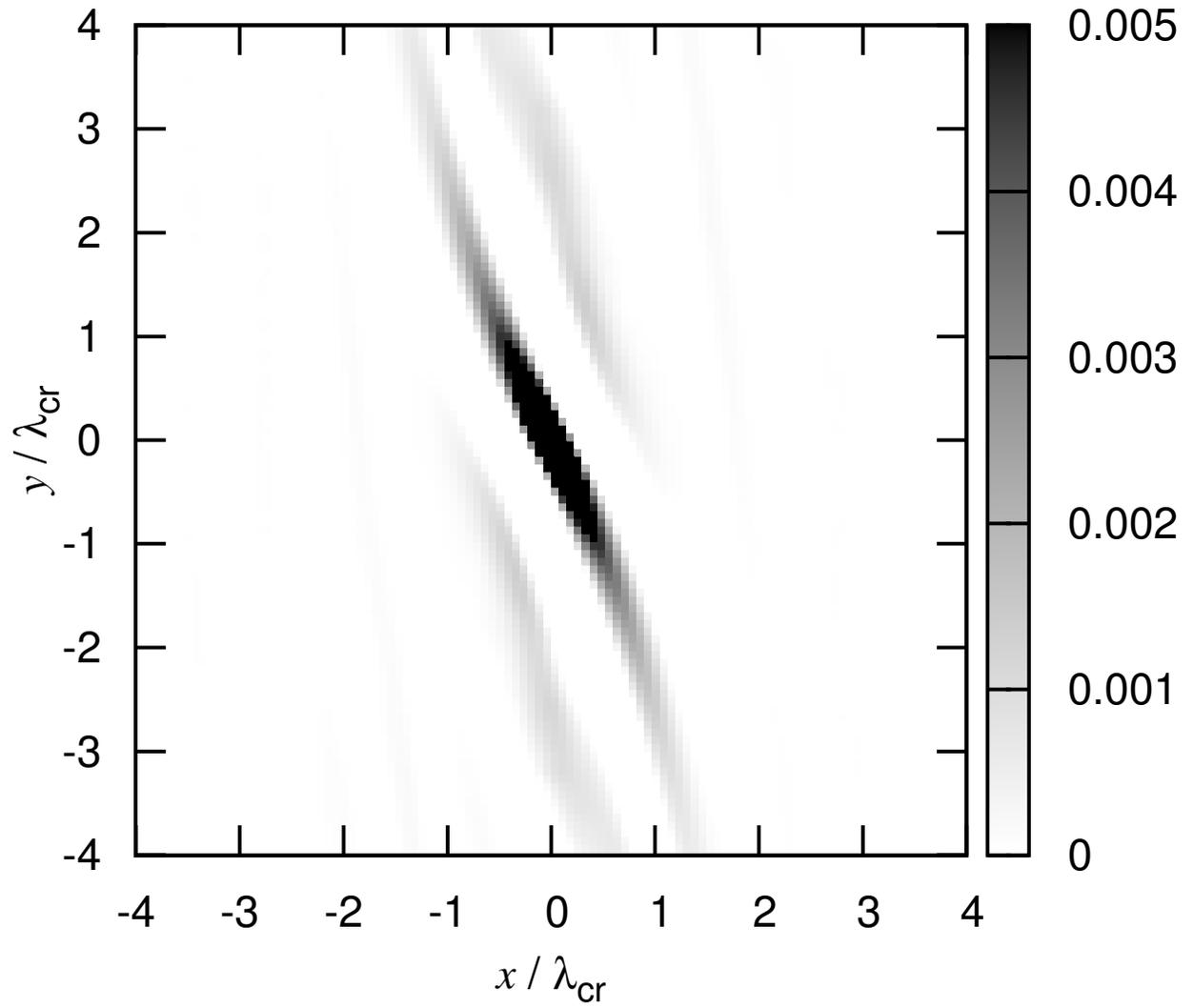}
  \end{center}
  \caption{Time averaged spatial autocorrelation function of the surface density $\xi$ for $\kappa/\Omega=1.4$ (model k4). }
  \label{fig:cor}
\end{figure}

\begin{figure}
  \begin{center}
	\plottwo{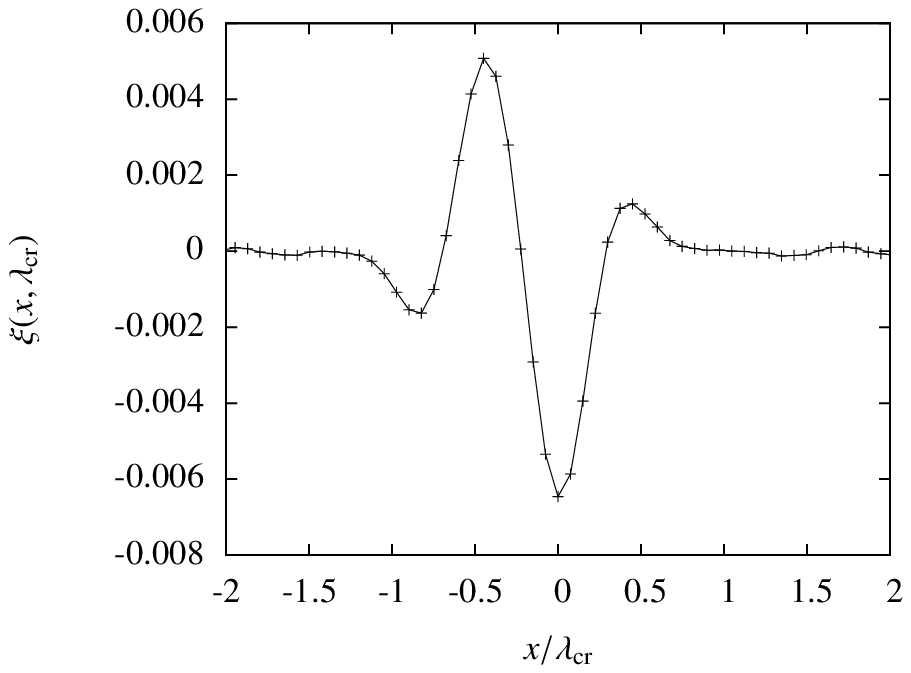}{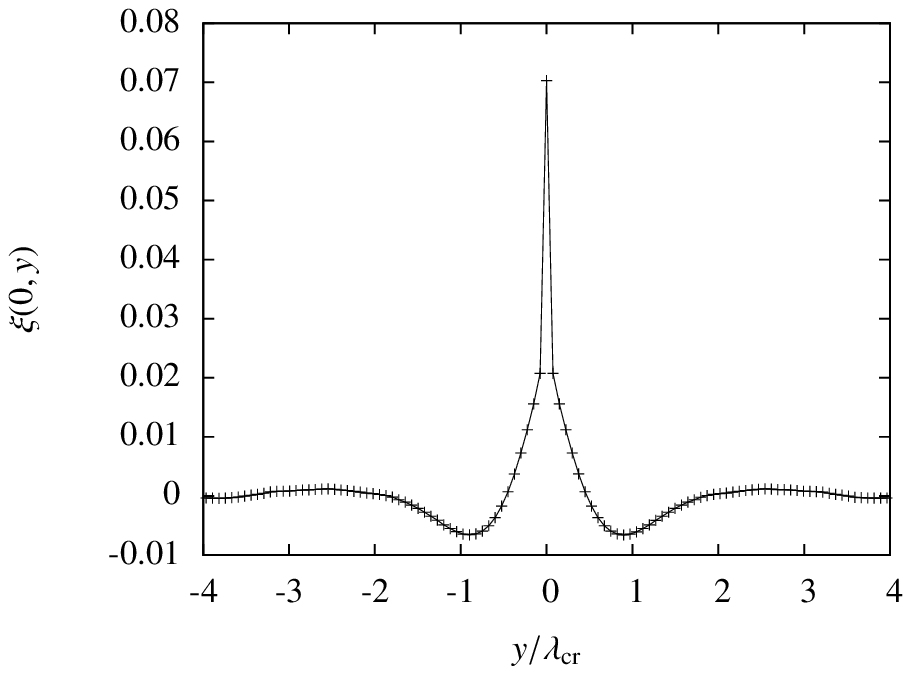}
  \end{center}
  \caption{Time averaged spatial autocorrelation function $\xi(x,\lambda_\mathrm{cr})$ (left) and $\xi(0,y)$ (right) for $\kappa/\Omega=1.4$ (model k4). }
  \label{fig:corxy}
\end{figure}

The Fourier transformation method with large wavelength that is comparable to the box size would be inaccurate.
Thus, we also  adopt another independent method to evaluate the dominant wave.
We introduce the spatial autocorrelation function of the surface density (Paper I)
\begin{equation}
	\xi(x,y) = -1 + \frac{1}{\Sigma_0^2 L^2} \int \!\!\! \int _{-L/2}^{L/2} \Sigma(x+x', y+y') \Sigma(x',y') \mathrm{d}x' \mathrm{d}y'.
\end{equation}
We calculate the time average of the autocorrelation over $t \Omega = 1 \mbox{--} 5$.
Figure \ref{fig:cor} presents the spatial autocorrelation for model k4.
The clear inclined structure at the center shows the averaged structure of the spirals, which is trailing.
We also observe two faint structures that are almost parallel to the clear structure at the center.
These faint structures correspond to the neighbouring spirals.
Therefore, the separation between these structures is the wavelength of the dominant spiral.
Figure \ref{fig:corxy} presents $\xi(x,\lambda_\mathrm{cr})$ and $\xi(0,y)$, which show the damped oscillation. 
Note that in order to show the structure in the $x$ direction we use $\xi(x,\lambda_\mathrm{cr})$ instead of $\xi(x,0)$.
This is because the neighbouring spirals on $y=0$ is too faint to extract.
Thus we focus on the line $y= \lambda_\mathrm{cr}$ and use $\xi(x,\lambda_\mathrm{cr})$.
We calculate the wavelength from the separation between the largest and the second largest peaks and obtain $\tilde \lambda_{x,\mathrm{max}}=0.92$ and $\lamymax=2.57$.

The results for the other models are summarized in Table \ref{tbl:model1}.
Basically, the spatial correlation method gives the same results as those by the Fourier transformation method.
For models k0, the azimuthal wavelength by the Fourier transformation method is $\lamymax=5.00$, which is 13 \% larger than that by the spatial correlation method.
It is because the Fourier transformation method with large wavelength is inaccurate.
When the box size is $15$, the wavelength that can be treated in the Fourier transformation is $15, 7.5, 5, 3.75, \cdots$. 
Therefore, for model k0, the spatial correlation method is more suitable.
For models k7 and k8, there is a large difference in $\tilde \lambda_{x,\mathrm{max}}$ and $\lamymax$ between these two methods.
For the large $\kappa/\Omega$ models, the amplitude of spiral arms is small. 
Thus, the noise may affect the estimates.

\subsection{Comparison with the Julian-Toomre Model}

\begin{figure}
  \begin{center}
  	\plotone{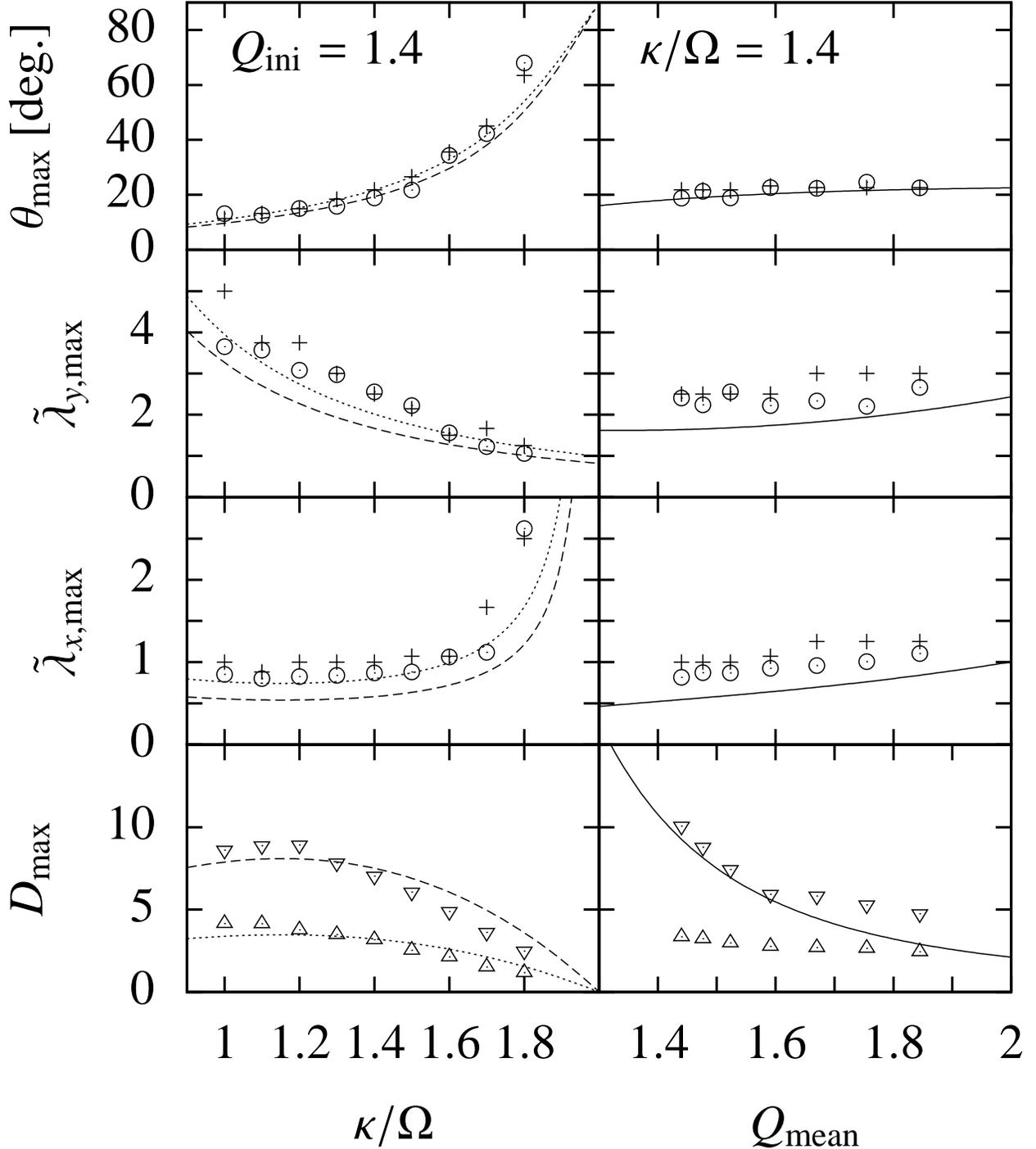}
  \end{center}
  \caption{(Left panel) Pitch angle $\theta_\mathrm{max}$, the azimuthal wavelength $\lamymax$, the radial wavelength $\tilde \lambda_{x,\mathrm{max}}$, and the amplification factor $D_\mathrm{max}$ calculated by Fourier transformation (plus) and autocorrelation function (circle) from top to bottom, respectively.
	The dashed and dotted curves denote the Julian-Toomre model fitting formulae given by Equations (\ref{eq:fit_pitch2}), (\ref{eq:fit_xf2}), (\ref{eq:fit_xl}), and (\ref{eq:fit_amp}) for $Q=1.5$ (Dashed) and $1.8$ (dotted).
	In the bottom panel, triangles up and down denote the amplification factor $D_0$ and $D_3$, respectively.
  (Right panel) Same as the left panel, but it shows the dependence on $Q_\mathrm{mean}$ with $\kappa/\Omega=1.4$. The solid curve shows the fitting formulae for $\kappa/\Omega=1.4$.}
  \label{fig:dep_kappa}

\end{figure}

The left panel of Figure \ref{fig:dep_kappa} shows the dependence of $\theta_\mathrm{max}$, $\lamymax$, $\tilde \lambda_{x,\mathrm{max}}$, and $D_\mathrm{max}$ on $\kappa/\Omega$. 
The pitch angle increases with $\kappa/\Omega$ as shown in Paper I.
The pitch angle by the Julian-Toomre model Equation (\ref{eq:fit_pitch2}) excellently agrees with that by the $N$-body simulations for $\kappa/\Omega < 1.7$.
The pitch angle by the $N$-body simulations is slightly larger than Equation (\ref{eq:fit_pitch2}) for  $\kappa/\Omega \ge 1.7$.

The general trend of the dependence of $\lamymax$ on $\kappa/\Omega$ in the $N$-body simulations is quite similar to that in the Julian-Toomre model, Equation (\ref{eq:fit_xf2}).
The azimuthal wavelength $\lamymax$ decreases with $\kappa/\Omega$.
For $\kappa/\Omega < 1.6$, $\lamymax$ in the $N$-body simulations is slightly larger than Equation (\ref{eq:fit_xf2}), but its difference is small.
The cause of the difference is unclear.  
One of the possible reasons is the nonlinear effect.  
Since the amplitude is large for small $\kappa/\Omega$, the nonlinearity may be important. 
Another reason is the breakdown of the assumption used to derive the fitting formulae.
We assumed that the observed spirals correspond to the wave with the maximum amplification factor.
This is justified when all the leading waves have the same amplitude.
However, in reality, it is not the case, thus the wave with the maximum amplification factor may not correspond to the spirals in a strict sense. 
For $\kappa/\Omega \ge 1.6$, the $N$-body simulations perfectly agree with Equation (\ref{eq:fit_xf2}).

Both the Julian-Toomre model Equation (\ref{eq:fit_xl}) and the $N$-body simulations show that $\tilde \lambda_{x,\mathrm{max}}$ is almost constant for $\kappa/\Omega < 1.7$ and it increases abruptly for $\kappa/\Omega \ge 1.7$.
For all $\kappa/\Omega$, $\tilde \lambda_{x,\mathrm{max}}$ in the $N$-body simulations is slightly larger than Equation (\ref{eq:fit_xl}).

\begin{figure}
  \begin{center}
  	\plotone{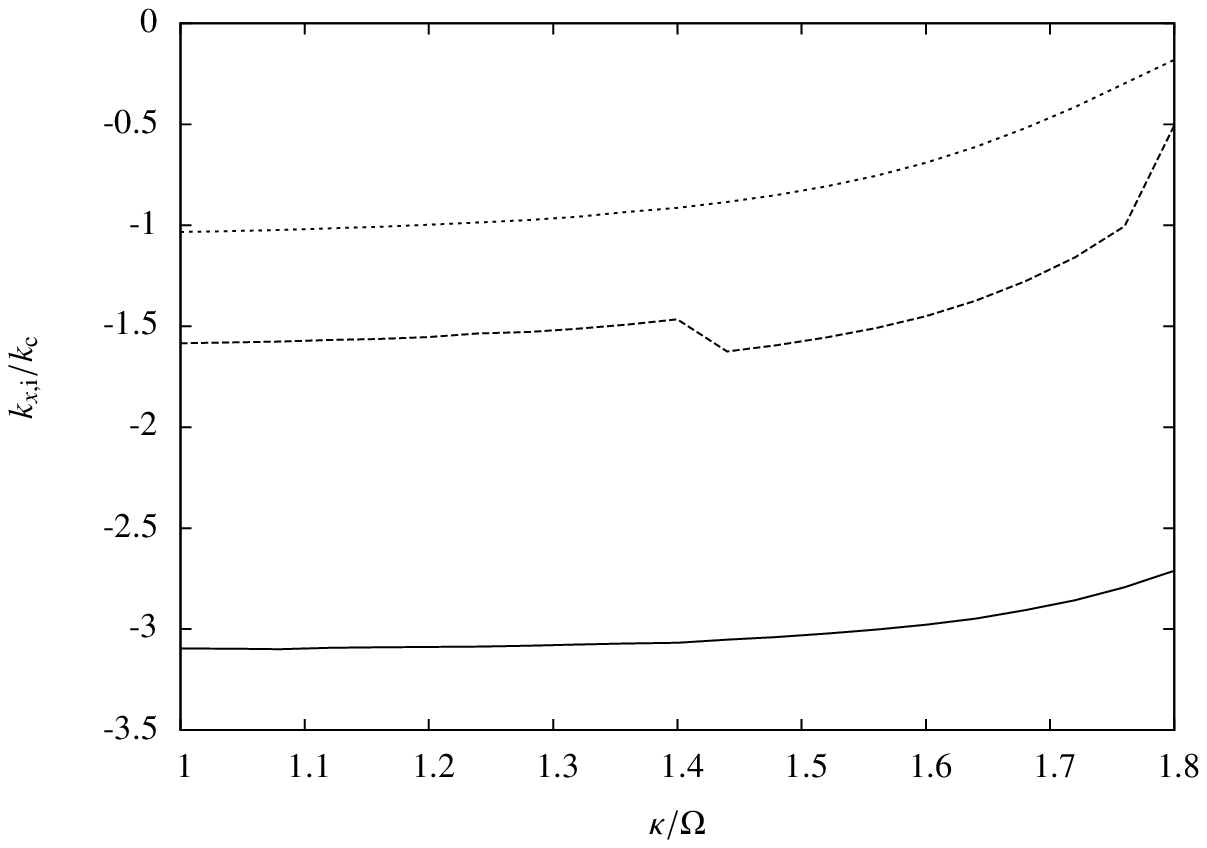}
  \end{center}
  \caption{Seed wavenumber $k_{x,\mathrm{i}}$ estimated by the linear analysis as a function of $\kappa / \Omega$ for $Q=1.0$ (solid curve), $1.4$ (dashed curve), and $1.8$ (short dashed curve). 
}
  \label{fig:dep_kappa_kx}

\end{figure}

In $N$-body simulations it is difficult to define the amplification factor in a strict sense because we cannot extract the corresponding initial amplitude.
Thus, we introduce an alternative quantity of the amplification.
First we find the maximum Fourier amplitude $a_\mathrm{max}$ and the corresponding wavelengths $k_{x,\mathrm{max}}$ and $k_{y,\mathrm{max}}$.
Next, since $k_y$ remains constant in the linear theory, we define the seed amplitude $a_\mathrm{seed}(C)$ as the Fourier amplitude 
with the wavenumbers $k_x= - C k_\mathrm{cr}$ and $k_y = k_{y,\mathrm{max}}$.
In the linear analyses, the wavenumber $k_x$ for the seed amplitude is calculated by $k_{x,\mathrm{i}} = 2 A k_{y,\mathrm{max}} t_\mathrm{i}$ for the most amplified wave.
Figure \ref{fig:dep_kappa_kx} shows $k_{x,\mathrm{i}}$ and we find that $ 0<C< 3$ is sufficient for $1.0 \le \kappa/\Omega \le 1.8$ and $1.0 \le Q \le 1.8$.
We define the amplification factor as $D_C= a_\mathrm{max}/ a_\mathrm{seed}(C)$.
For $C=0$, this ratio $D_0$ corresponds to the amplification factor of a half-swing \citep{Toomre1981}.
From the simulations we found that $a_\mathrm{seed}(C)$ decreases with increasing $C$. 
Therefore it is expected that $D_0$ corresponds to the lower bound of the amplification factor.
On the other hand, if we adopt $C=3$, $D_3$ corresponds to the upper bound.
Note that we cannot directly compare $D_0$ and $D_3$ with the amplification factor $D_\mathrm{max}$ because its definition is different.
We do not discuss the absolute value of $D$ but focus on the dependence of $D$ on $\kappa/\Omega$ and $Q$.
The Julian-Toomre model Equation (\ref{eq:fit_amp}) shows that $D_\mathrm{max}$ decreases with $\kappa/\Omega$ for $Q \ge 1.6$.
The $N$-body simulations show the similar trend.
In the $N$-body simulations $D_0$ and $D_3$ monotonically decreases with $\kappa/\Omega$.

The right panel of Figure \ref{fig:dep_kappa} shows the dependence of these parameters on $Q$ for $\kappa/\Omega=1.4$. 
Since $Q$ is not steady but increases with time, we used the time averaged $Q$ over $t \Omega = 1 \mbox{--} 5$, which is summarized as $Q_\mathrm{mean}$ in Table \ref{tbl:model1}.
We exclude the models with $Q_\mathrm{ini}=1.0$ and $1.1$ because the change of $Q$ is significant.
In the $N$-body simulations, the pitch angle is almost independent of $Q_\mathrm{mean}$ and agrees well with those by the linear theory.
The linear theory predicts that the wavelengths $\lamymax$ and $\tilde \lambda_{x,\mathrm{max}}$ increase with $Q$ for $Q>1.4$.
It seems that $\lamymax$ measured by the Fourier transformation increases with $Q$ but $\lamymax$ measured by the autocorrelation function is almost constant.
The wavelength $\lamymax$ is about $0.5 \lambda_\mathrm{cr}$ longer than those by the linear theory.
In the $N$-body simulations, $\tilde \lambda_{x,\mathrm{max}}$ increases with $Q_\mathrm{mean}$. The tendency is consistent with the linear theory.
Since the dependence on $Q$ is weak and the range of $Q$ in the $N$-body simulations is narrow, 
it is difficult to confirm if the dependence of the wavelengths on $Q$ by $N$-body simulations agrees with those by the linear theory quantitatively. 
However at least we find that the tendency is generally the same.
In the $N$-body simulations, $D_3$ decreases with $Q_\mathrm{mean}$, which is consistent with the linear theory.
The half-swing amplification factor $D_0$ slightly decreases but is almost constant, which shows that $D_0$ may not be a good indicator for the degree of the amplification for small $Q$.

\begin{deluxetable}{ccccccccccccc}
  \tabletypesize{\scriptsize}
  \tablecaption{Model Parameters and Results\label{tbl:model1}}
  \tablewidth{0pt}
  \tablehead{ \colhead{}
	& \colhead{}
	& \colhead{}
	& \colhead{}
	& \colhead{}
	& \multicolumn{3}{c}{Fourier Transformation} 
	& \multicolumn{3}{c}{Autocorrelation Function} \\
     \colhead{Model} 
	& \colhead{$\kappa / \Omega$}
	& \colhead{$\Gamma$}
	 & \colhead{$Q_\mathrm{ini}$}
	 & \colhead{$Q_\mathrm{mean}$}
	 & \colhead{$\tilde \lambda_{x,\mathrm{max}}$}
	& \colhead{$\lamymax$}
	& \colhead{$\theta_{\mathrm{max}}$}
	 & \colhead{$\tilde \lambda_{x,\mathrm{max}}$}
	& \colhead{$\lamymax$}
	& \colhead{$\theta_{\mathrm{max}}$}
  }
  \startdata
	k0 & 1.0 & 1.50 & 1.4 & 1.64 & 1.00 & 5.00 & 11.3 & 0.85 & 3.65 & 13.1 \\
	k1 & 1.1 & 1.40 & 1.4 & 1.61 & 0.88 & 3.75 & 13.2 & 0.80 & 3.57 & 12.6 \\
	k2 & 1.2 & 1.28 & 1.4 & 1.58 & 1.00 & 3.75 & 14.9 & 0.82 & 3.08 & 15.0 \\
	k3 & 1.3 & 1.15 & 1.4 & 1.56 & 1.00 & 3.00 & 18.4 & 0.84 & 2.97 & 15.8 \\
	k4 & 1.4 & 1.02 & 1.4 & 1.52 & 1.00 & 2.50 & 21.8 & 0.87 & 2.55 & 18.8 \\
	k5 & 1.5 & 0.88 & 1.4 & 1.49 & 1.07 & 2.14 & 26.6 & 0.88 & 2.22 & 21.7 \\
	k6 & 1.6 & 0.72 & 1.4 & 1.46 & 1.07 & 1.50 & 35.5 & 1.06 & 1.56 & 34.3 \\
	k7 & 1.7 & 0.56 & 1.4 & 1.43 & 1.67 & 1.67 & 45.0 & 1.12 & 1.23 & 42.3 \\
	k8 & 1.8 & 0.38 & 1.4 & 1.42 & 2.50 & 1.25 & 63.4 & 2.62 & 1.06 & 68.0 \\
  	\\
	q2 & 1.4 & 1.02 & 1.2 & 1.44 & 1.00 & 2.50 & 21.8 & 0.82 & 2.41 & 18.7 \\
	q3 & 1.4 & 1.02 & 1.3 & 1.48 & 1.00 & 2.50 & 21.8 & 0.87 & 2.24 & 21.3 \\
	q5 & 1.4 & 1.02 & 1.5 & 1.59 & 1.07 & 2.50 & 23.2 & 0.93 & 2.22 & 22.6 \\
	q6 & 1.4 & 1.02 & 1.6 & 1.67 & 1.25 & 3.00 & 22.6 & 0.96 & 2.33 & 22.3 \\
	q7 & 1.4 & 1.02 & 1.7 & 1.75 & 1.25 & 3.00 & 22.6 & 1.01 & 2.20 & 24.5 \\
	q8 & 1.4 & 1.02 & 1.8 & 1.84 & 1.25 & 3.00 & 22.6 & 1.11 & 2.66 & 22.5 \\
\enddata
\end{deluxetable}

\section{Summary and Discussion}
\subsection{Summary}
We investigated the swing amplification based on the Julian-Toomre model \citep{Julian1966}.
Optimizing the azimuthal wavelength $\lamy$ and the initial time $t_\mathrm{i}$ that corresponds to the initial radial wavelength, we calculated the maximum amplification factor $D_\mathrm{max}$, and the corresponding wavelengths $\lamymax$, and  $\tilde \lambda_{x,\mathrm{max}}$ and pitch angle $\theta_\mathrm{max}$.
As already shown in \cite{Michikoshi2014} (Paper I), $\theta_\mathrm{max}$ is in good agreement with $N$-body simulations.
We investigated the dependencies of $\theta_\mathrm{max}$, $\lamymax$, $\tilde \lambda_{x,\mathrm{max}}$, and $D_\mathrm{max}$ on $\kappa/\Omega$ and Toomre's $Q$.
The azimuthal wavelength $\lamymax$, that is $X$ parameter in \cite{Julian1966}, decreases with $\kappa/\Omega$ from $4$ to $1$, which is consistent with the previous works \citep{Toomre1981, Athanassoula1984, Dobbs2014}.
We found that $\lambda_{x,\mathrm{max}}$ is almost constant and slightly smaller than $\lambda_\mathrm{cr}$ for $\kappa/\Omega<1.5$.
For $\kappa/\Omega>1.5$, $\lambda_{x,\mathrm{max}}$ increases with $\kappa/\Omega$.
The amplification factor $D_\mathrm{max}$ decreases with $Q$ sensitively.

Next, we performed local $N$-body simulations and calculated $\theta_\mathrm{max}$, $\lamymax$, $\tilde \lambda_{x,\mathrm{max}}$, and $D_\mathrm{max}$.
We extracted the most amplified wave from $N$-body simulations by utilizing the Fourier transformation and the autocorrelation function of the surface density.
We found that the dependencies of $\theta_\mathrm{max}$, $\lamymax$, and $\tilde \lambda_{x,\mathrm{max}}$ on $\kappa/\Omega$ agree well with our fitting formulae derived from the Julian-Toomre model.
The dependence of $\theta_\mathrm{max}$ on $\kappa/\Omega$ (which is a function of the shear rate, $\Gamma$, see Eq. (\ref{eq:gamma})) was already confirmed in Paper 1, but the formulae of $\lamymax$ was newly derived and confirmed with $N$-body simulations.
We also examine the dependencies of these parameters on $Q$.
Because the dependence on $Q$ is weak and it is difficult to control $Q$ in $N$-body simulations, we could not clearly confirm the $Q$ dependence
of these parameters completely but at least the fitting formula and $N$-body simulations are generally consistent.
This could be tested with more controlled $N$-body simulations in the future study.

The overall activity of spiral arms cannot be explained only by the linear theory of the swing amplification.
The nonlinear processes are important \citep{Fuchs2005, DOnghia2013}.
However, in Paper I and this paper, we showed that the linear theory of the swing amplification explains the spiral arm structures quantitatively. 
The linear theory elucidates some important aspects of the basic physics of spiral arm dynamics.

\subsection{Number of Spiral Arms}
We show an example of applications of the fitting formula to an observation.
We can estimate the number of spiral arms for multi-arm spiral galaxies from $\lamymax$ \citep{Carlberg1985a}.
The number of spiral arms is estimated by 
\begin{equation}
  m \simeq \frac{2 \pi R}{\lambda_y} = \frac{\kappa^2 R}{2 \pi G \Sigma_0 \lamymax},
  \label{eq:marm}
\end{equation}
where $R$ is the disk radius.
In the previous works \citep[e.g.,][]{Carlberg1985a, Fujii2011}, $\lamymax=2$ was assumed, which is valid for $\Gamma \simeq 1$.
To obtain a more general formula, we substitute Equation (\ref{eq:fit_xf_simp}) into Equation (\ref{eq:marm}), 
where we approximate the orbital frequency as
\begin{equation}
  \Omega^2 \simeq \frac{G M_\mathrm{tot}}{R^3} \simeq \frac{\pi G \Sigma_\mathrm{0}}{R f_\mathrm{disk}}, \label{eq:omega}
\end{equation}
where $M_\mathrm{tot}$ is the total mass including the dark halo and  $f_\mathrm{disk}$ is the ratio of disk mass to total mass.
The ratio of disk mass to halo mass $\zeta$ is also often used, which has the relation $f_\mathrm{disk}=\zeta/(1+\zeta)$.
Thus, we obtain the number of spiral arms
\begin{equation}
  m \simeq 0.230 \frac{\kappa^4}{f_\mathrm{disk} Q \Omega^4} = 0.922 \frac{(2-\Gamma)^2}{f_\mathrm{disk} Q}.
  \label{eq:numarm}
\end{equation}
As already shown by \cite{Carlberg1985a}, the number of spiral arms is in inversely proportion to $f_\mathrm{disk}$.
The high resolution $N$-body simulations support this tendency \citep{Fujii2011, DOnghia2013}.
This estimate also shows that $m$ decreases with $\Gamma$, which has not been clearly confirmed by $N$-body simulations yet.

We can rewrite Equation (\ref{eq:numarm}) by using $\theta$.
Eliminating $\kappa$ from Equations (\ref{eq:fit_pitch}) and (\ref{eq:numarm}), we obtain
\begin{equation}
  m \simeq \frac{0.230}{Q f_\mathrm{disk}} \left( \frac{14 \tan \theta}{1+\sqrt{1+49 \tan^2 \theta}} \right)^{4},
\end{equation}
which can be reduced to 
\begin{eqnarray}
m \simeq \left\{ \begin{array}{ll}
	\displaystyle{\left( 4.0 \left(\frac{\theta}{20^\circ} \right) - 1.36 \right)\left(\frac{f_\mathrm{disk}}{0.2}\right)^{-1}  \left(\frac{Q}{1.5}\right)^{-1}} & (10 ^\circ < \theta < 30 ^\circ) \\
	\displaystyle{6.1 \left(\frac{\theta }{40^\circ} \right)\left(\frac{f_\mathrm{disk}}{0.2}\right)^{-1}  \left(\frac{Q}{1.5}\right)^{-1}} & (30 ^\circ < \theta < 60 ^\circ) \\
  \end{array} \right.,
\end{eqnarray}
where the relative error is $\lesssim 5\%$.
The number of spiral arm $m$ increases with $\theta$. 
Note that since this estimate is based on the local approximation we should be careful when we apply it to the case for small $m$.

Note that the formula of the number of spiral arms presented here is based on the extremely simplified assumptions such as Equation (\ref{eq:omega}).
In fact, the number of spiral arms can depend on the galactocentric distance \citep{Bottema2003, DOnghia2015}.
For more realistic estimation, we have to consider the detailed disk and dark halo models.
\cite{DOnghia2015} considered the exponential stellar disk and the dark halo model of the Hernquist mass profile
and obtained that the number of spiral arms depends on the galactocentric distance by assuming that the azimuthal wavelength $X=1.5\mbox{--}2$ that is valid for $\Gamma \simeq 1$.
Using our azimuthal wavelength formula (Eq. (\ref{eq:fit_xf_simp})), we can obtain more general analytic formula that also depends on $\theta$ or $\Gamma$.

\acknowledgments{
  Numerical computations were carried out on the GRAPE system at Center for Computational Astrophysics, National Astronomical Observatory of Japan.
}


\begin{thebibliography}{27}
\expandafter\ifx\csname natexlab\endcsname\relax\def\natexlab#1{#1}\fi

\bibitem[{{Athanassoula}(1984)}]{Athanassoula1984}
{Athanassoula}, E. 1984, \physrep, 114, 319

\bibitem[Baba(2015)]{Baba2015} Baba, J.\ 2015, \mnras, 454, 2954 

\bibitem[{{Baba} {et~al.}(2009){Baba}, {Asaki}, {Makino}, {Miyoshi}, {Saitoh},
  \& {Wada}}]{Baba2009}
{Baba}, J., {Asaki}, Y., {Makino}, J., {Miyoshi}, M., {Saitoh}, T.~R., \&
  {Wada}, K. 2009, \apj, 706, 471

\bibitem[{{Baba} {et~al.}(2013){Baba}, {Saitoh}, \& {Wada}}]{Baba2013}
{Baba}, J., {Saitoh}, T.~R., \& {Wada}, K. 2013, \apj, 763, 46

\bibitem[{{Bottema}(2003)}]{Bottema2003}
{Bottema}, R. 2003, \mnras, 344, 358

\bibitem[{{Carlberg} \& {Freedman}(1985)}]{Carlberg1985a}
{Carlberg}, R.~G. \& {Freedman}, W.~L. 1985, \apj, 298, 486

\bibitem[{{Dobbs} \& {Baba}(2014)}]{Dobbs2014}
{Dobbs}, C. \& {Baba}, J. 2014, \pasa, 31, 35

\bibitem[{{D'Onghia}(2015)}]{DOnghia2015}
{D'Onghia}, E. 2015, \apjl, 808, L8

\bibitem[{{D'Onghia} {et~al.}(2013){D'Onghia}, {Vogelsberger}, \&
  {Hernquist}}]{DOnghia2013}
{D'Onghia}, E., {Vogelsberger}, M., \& {Hernquist}, L. 2013, \apj, 766, 34

\bibitem[{{Fuchs}(2001)}]{Fuchs2001}
{Fuchs}, B. 2001, \aap, 368, 107

\bibitem[{{Fuchs} {et~al.}(2005){Fuchs}, {Dettbarn}, \& {Tsuchiya}}]{Fuchs2005}
{Fuchs}, B., {Dettbarn}, C., \& {Tsuchiya}, T. 2005, \aap, 444, 1

\bibitem[{{Fujii} {et~al.}(2011){Fujii}, {Baba}, {Saitoh}, {Makino}, {Kokubo},
  \& {Wada}}]{Fujii2011}
{Fujii}, M.~S., {Baba}, J., {Saitoh}, T.~R., {Makino}, J., {Kokubo}, E., \&
  {Wada}, K. 2011, \apj, 730, 109

\bibitem[{{Goldreich} \& {Lynden-Bell}(1965)}]{Goldreich1965}
{Goldreich}, P. \& {Lynden-Bell}, D. 1965, \mnras, 130, 125

\bibitem[{{Grand} {et~al.}(2013){Grand}, {Kawata}, \& {Cropper}}]{Grand2013}
{Grand}, R.~J.~J., {Kawata}, D., \& {Cropper}, M. 2013, \aap, 553, A77

\bibitem[{{Julian} \& {Toomre}(1966)}]{Julian1966}
{Julian}, W.~H. \& {Toomre}, A. 1966, \apj, 146, 810

\bibitem[{Kawai \& Fukushige(2006)}]{Kawai2006}
Kawai, A. \& Fukushige, T. 2006, 48

\bibitem[{{Kokubo} \& {Ida}(1992)}]{Kokubo1992}
{Kokubo}, E. \& {Ida}, S. 1992, \pasj, 44, 601

\bibitem[{{Lin} \& {Shu}(1964)}]{Lin1964}
{Lin}, C.~C. \& {Shu}, F.~H. 1964, \apj, 140, 646

\bibitem[{{Lin} \& {Shu}(1966)}]{Lin1966}
---. 1966, Proceedings of the National Academy of Science, 55, 229

\bibitem[{{Michikoshi} \& {Kokubo}(2014)}]{Michikoshi2014}
{Michikoshi}, S. \& {Kokubo}, E. 2014, \apj, 787, 174 (Paper I)

\bibitem[{{Seigar} {et~al.}(2005){Seigar}, {Block}, {Puerari}, {Chorney}, \&
  {James}}]{Seigar2005}
{Seigar}, M.~S., {Block}, D.~L., {Puerari}, I., {Chorney}, N.~E., \& {James},
  P.~A. 2005, \mnras, 359, 1065

\bibitem[{{Seigar} {et~al.}(2006){Seigar}, {Bullock}, {Barth}, \&
  {Ho}}]{Seigar2006}
{Seigar}, M.~S., {Bullock}, J.~S., {Barth}, A.~J., \& {Ho}, L.~C. 2006, \apj,
  645, 1012

\bibitem[{{Sellwood}(2000)}]{Sellwood2000}
{Sellwood}, J.~A. 2000, \apss, 272, 31

\bibitem[{{Sellwood} \& {Carlberg}(1984)}]{Sellwood1984}
{Sellwood}, J.~A. \& {Carlberg}, R.~G. 1984, \apj, 282, 61

\bibitem[{{Toomre}(1964)}]{Toomre1964}
{Toomre}, A. 1964, \apj, 139, 1217

\bibitem[{{Toomre}(1981)}]{Toomre1981}
  ---. 1981, Structure and Evolution of Normal Galaxies (Cambridge: Cambridge Univ. Press), 111

\bibitem[{{Toomre} \& {Kalnajs}(1991)}]{Toomre1991}
{Toomre}, A. \& {Kalnajs}, A.~J. 1991, 341

\bibitem[{{Wisdom} \& {Tremaine}(1988)}]{Wisdom1988}
{Wisdom}, J. \& {Tremaine}, S. 1988, \aj, 95, 925

\end{thebibliography}
\end{document}